\documentclass[12pt]{article}
\usepackage{graphicx}
 \usepackage{amssymb}
 \usepackage{amsmath}
\usepackage{cite}
 \usepackage{geometry}               
\def\gsim{\mathrel{\rlap{\lower4pt\hbox{\hskip1pt$\sim$}}
    \raise1pt\hbox{$>$}}}
    \def\lsim{\mathrel{\rlap{\lower4pt\hbox{\hskip1pt$\sim$}}
    \raise1pt\hbox{$<$}}}

\newcommand {\slsh} [1] {\not{\hbox{\kern-4pt${#1}$}}}

\newcommand {\beq} {\begin{equation}}
\newcommand {\eeq} {\end{equation}}
  \newcommand {\ber}{\begin{eqnarray*}}
  \newcommand {\eer} {\end{eqnarray*}}
\newcommand {\beqn}{\begin{eqnarray}}
  \newcommand {\eeqn} {\end{eqnarray}}

\begin{document}
\begin{titlepage}
\begin{flushright}
{FTPI-MINN-05/44\\
UMN-TH-2417/05\\
October 13/05
}
\end{flushright}
\vskip 0.8cm

\centerline{{\Large \bf $k$ Strings from Various Perspectives:}}

\vskip 0.2cm
\centerline{\Large \bf
QCD, Lattices, String Theory
and Toy Models\,\footnote{Based on invited 
talks delivered at 
the Workshop {\sl  Understanding Confinement/Zakharov-Fest}, Ringberg Castle, Tegernsee, May 16-21, Germany, and  the  
{\sl Cracow School of Theoretical Physics},
Zakopane, Poland, June 3-12, 2005.}}

\vskip 1cm
\centerline{\large  M. Shifman  
}
\vskip 0.1cm
\begin{center}
{\em  William I. Fine Theoretical Physics Institute, 
University
of Minnesota, Minneapolis, MN 55455, USA }
\end{center}

\vspace{1.5cm}

\begin{abstract}

\vspace{.3cm}

I review the status of  the issue of the $k$-string tension in
Yang--Mills theory. After a summary of known facts
I discuss a weakly coupled four-dimensional Yang--Mills theory
that supports non-Abelian strings and can, in certain aspects,  serve as a toy model 
for QCD strings. In the second part of the talk I present original results obtained 
in a two-dimensional toy model which provides some evidence for the sine formula.

 \end{abstract}

\end{titlepage}

\section{Preamble}
\label{fi}
\renewcommand{\theequation}{\thesection.\arabic{equation}}
\setcounter{equation}{0}

Valya Zakharov was the first to introduce me to QCD,
even before QCD was officially born. In the summer of 1972 he was sent, as the ITEP Ambassador, to the Rochester conference which  took place at Fermilab that year. After his return, he told us of his impressions.

Apparently, Gell-Mann's talk produced the strongest impression on Valya
since he kept saying that Gell-Mann had been preaching octet gluons
as  mediators of the inter-quark force, and we ought to do something.
I was in the very beginning of my PhD work at that time, and knew very little
as to how to orient myself in the sea of literature, and whom to trust.
Valya repeated, more than once, that Gell-Mann had a direct line to god
--- Gell-Mann's revelations ought to be taken seriously. 

It would be fair to
say that Valya's persistence and foresight shaped my career
to a large extent. 
It is gratifying to note that now, 33 years later, he continues  
a noble mission of analytic thinking,  deep insight and promotion of innovative ideas in the lattice QCD community. Thank you, Valya,
and Happy Birthday ...

\vspace{7mm}

{\huge Part I: Review}

\section{Introduction}
\label{si}
\renewcommand{\theequation}{\thesection.\arabic{equation}}
\setcounter{equation}{0}

For this talk I chose a topic  seemingly fitting well the 
scope of Zakharov's interests which in the recent years revolve
around various mechanisms of color confinement in Yang--Mills (YM) theories.
The issue I want to address is  the $k$-string tension.
Significant effort has been invested recently in QCD and YM theories at large
in the studies of  flux tubes induced by color sources
in higher representations of SU($N$), mostly in connection
--- but not exclusively --- with high-precision lattice calculations
(for a review see \cite{grin} and references therein; see also Sect.~\ref{ndol}). 
If a source has $k$ fundamental color indices ($N$-ality $k$),
the flux tube it generates is referred to as the $k$-string. The question
of $k$ and $N$ dependence of the $k$-string tension
is one of the central questions of color-confining dynamics.

For many years the prevailing hypothesis
was that of the so-called Casimir scaling, the genesis of which  can 
be seemingly traced back to various models based on one-gluon exchange,
popular in the 1970's and 80's,
as well as to the strong-coupling lattice expansions.
Surprisingly, only recently it was 
realized \cite{Armoni-one,Armoni-two} that the Casimir scaling is in 
direct contradiction with the  $1/N$ expansion
in YM theory.

Meanwhile, an alternative construction emerged
which does have an appropriate $1/N$ expansion. It goes under the
name of the sine formula for the $k$-string tension.
Originally it was suggested by Douglas and 
Shenker \cite{Douglas:1995nw} in connection with
${\cal N}=2$ super-Yang--Mills model.
Arguments in favor of this formula
were obtained in MQCD and supersymmetric theories
(for a detailed discussion and a representative list
of references see \cite{Armoni-one,Armoni-two}). 

In spite of the abundance of arguments,
and the correct $1/N$ behavior inherent to the sine formula,
it has never been proven, and the issue of its relevance to QCD
remains open. This talk consists of two parts. Part I
is designed as a brief review of general ideas regarding QCD $k$-strings,
with the emphasis on developments after 2003. For a review of the pre-2003 situation
the reader is referred to \cite{Armoni-two}.
Part II is original. It presents a toy model 
in which the exact sine formula for an analog of the $k$-string tension
emerges in a natural way. This toy model is two-dimensional but it is
closely (in fact, ``genetically") related to a four-dimensional 
weakly coupled gauge model which has been recently developed, see 
\cite{ShifmanYung,GSY}, and references therein, and Sect.~\ref{naswc}.
Its remarkable feature is that it supports {\em non-Abelian} strings,
pretty close relatives of QCD strings, the main subject of our analysis.

%

%

%

\section{Strings/flux tubes in known phenomena}
\label{sft}
\renewcommand{\theequation}{\thesection.\arabic{equation}}
\setcounter{equation}{0}

A physical phenomenon
where flux tubes with well-studied properties 
are proven to play a crucial role is known from  1930s and to theorists
from  1950s. In 1957 Abrikosov published the paper
\cite{abri} entitled {\sl On the Magnetic Properties of Superconductors
of the Second Type} which deals with penetration of magnetic fields in bulk superconductors. Magnetic flux is conserved. If one places
a large bulk superconductor between two poles of a magnet, the magnetic field
must go through, but it cannot go through without destroying superconductivity.
Treating the Cooper pair condensation in the framework of the
Ginzburg--Landau theory, Abrikosov found
vortex-type solutions describing magnetic fields squeezed into thin tubes
carrying the total magnetic flux which is quantized.
The corresponding physical phenomenon is called the Meissner effect.\footnote{Walter Meissner and Robert Ochsenfeld discovered in 1933  that superconducting materials repelled magnetic fields. This effect is quite spectacular:
 magnets can   levitate  above superconducting materials.}
Superconductivity 
is destroyed in the
core of the tubes. The energy of such configurations is $\sigma L$
where $\sigma$ is the string tension and $L$ is the size of the 
 bulk superconductor pierced by the flux tube. The energy scales linearly with
the size. The flux tubes end on the poles of the magnet or, if 
magnetic monopole existed, they could end on the magnetic monopoles. 

The Abrikosov flux tubes are currently known as
Abelian or U(1) flux tubes. They emerge in  the Abelian Higgs model
(see \cite{NO}) due to the fact that $\pi_1 ({\rm U}(1))$ is non-trivial.
The gauge U(1) symmetry is spontaneously broken in the vacuum by
a condensate of a charged scalar field $\phi$, the Higgs field,  
which can be thought of as
representing the Cooper pair density. The string configuration, with a winding phase
of the scalar field $\phi$, is topologically stable. At large distances from string's core
$|\phi |$ coincides with its vacuum value, while inside the core
$\phi \to 0$. Thus, superconductivity is destroyed in string's core.
The magnetic flux transmitted through the
Abrikosov--Nielsen--Olesen (ANO) flux tube can be arbitrary integer number
(in appropriate units).

\section{Color confinement: dual Meissner effect hypothesis}
\label{ccd}
\renewcommand{\theequation}{\thesection.\arabic{equation}}
\setcounter{equation}{0}

Magnetic charges attached to the endpoints of the Abrikosov string 
are confined:
taking them apart would require infinite energy. 
In QCD we want chromoelectric charges, rather than chromomagnetic
ones, to be  attached to the endpoints of the chromoelectric
flux tubes and thus confined. 
Chromomagnetic charges must condense.
In the 1970s  't Hooft \cite{thooft}
and Mandelstam \cite{mandelstam}
put forward the hypothesis of a dual Meissner effect 
to explain color confinement in non-Abelian gauge
theories. Dual means that we take the Abrikosov theory and
replace everything electric by magnetic and vice versa.

The  't Hooft-Mandelstam hypothesis was formulated at a qualitative
level.  Since then people kept trying to find 
a quantitative framework 
in which one could  demonstrate
the occurrence of the dual Meissner effect in 
a controllable approximation:
formation of  chromoelectric flux tubes with properties compatible with
general ideas and existing data on color confinement.
This task proved to be extremely difficult.

A breakthrough achievement was the Seiberg--Witten
solution \cite{sw} of ${\cal N}=2$ supersymmetric 
Yang--Mills theory. They found massless monopoles 
at a certain point in the moduli space and,
adding a small $({\cal N}=2)$-breaking deformation,
demonstrated that they condense creating strings carrying 
a chromoelectric flux. This was the first quantitative implementation 
of the 't Hooft--Mandelstam hypothesis,
16 years after its inception! Needless to say, the
Seiberg--Witten result was
 met with great
enthusiasm in the community.

A more careful examination showed, however, that details of
the Seiberg--Witten confinement were quite different from 
those we expect in QCD-like theories. Indeed, a crucial aspect
of Ref.~\cite{sw} is  that the SU($N$)
gauge symmetry is first broken, at a high scale, down to U(1)$^{N-1}$,
which is then completely broken, at a much lower scale
where monopoles condense. Correspondingly,
the Seiberg-Witten model  strings are, in fact, the Abelian strings \cite{abri,NO}
of the Abrikosov--Nielsen--Olesen
type. This  leads  to an ``Abelian" confinement
whose structure does not resemble at all that of QCD. In particular, the
``hadronic" spectrum is much richer than that in QCD
\cite{Douglas:1995nw,matt}. For a review see \cite{MattS}.

Thus, the task of constructing and studying non-Abelian strings
in QCD and QCD-like theories
still stands. I will summarize recent (modest)  progress 
\cite{recent,Hanany,Auzzi,ShifmanYung,GSY,Tong,HananyTong,Markov} at the end of Part I. However, to begin with, it is instructive to discuss
what we know of QCD strings from  $1/N$, lattices, and other ideas. 

\section{QCD strings}
\label{qcds}
\renewcommand{\theequation}{\thesection.\arabic{equation}}
\setcounter{equation}{0}

The simplest QCD string is the flux tube that connects heavy (probe) color sources
in the fundamental representation. It is referred to as the fundamental string.
The fundamental string tension is of the order of
$\Lambda^2$ where $\Lambda$
is the dynamical scale parameter. Its transverse size is of the order of $\Lambda^{-1}$.
Both parameters are independent of the number of colors,
 and, besides $\Lambda$, can contain only
numerical factors.
In what follows I assume the gauge
group  to be SU($N$), with $N$, the number of colors, being a  free parameter and $g^2N$ fixed.

\begin{figure}
\centerline{\includegraphics[width=2.5in]{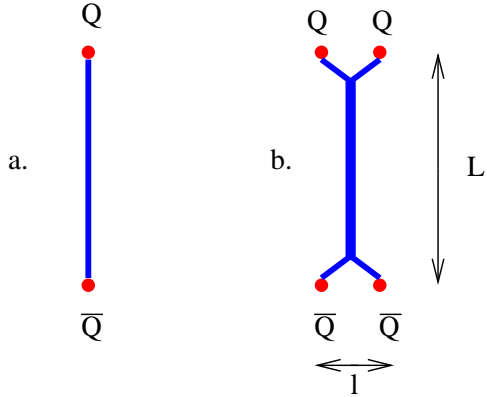}}
\caption{A flux tube for $k$-strings: (a) the
 fundamental  tube; (b) the 2-string tube.}
\label{st1}
\end{figure}

The  flux tubes attached to color sources
in higher representations of SU($N$) are known as $k$-strings, 
where $k$ denotes the
$N$-ality of the color representation under consideration.
The $N$-ality of the representation with $\ell$ upper and
$m$ lower indices (i.e. $\ell$ fundamental
and $m$ anti-fundamental)  is defined as 
\beq
k= |\ell -m|\,.
\eeq
 Figure  \ref{st1}
displays the fundamental string and 2-string. Since $N$ fundamental quarks
can form a color-singlet object (baryon), $N$-ality is defined mod $N$. 

$N$ fundamental strings collected together, as in Fig.~\ref {st1}b, will pass 
into a no-string state. This feature of non-Abelian strings critically distinguishes them
from the ANO strings. For the same reason, if, say, 
$N$ is odd, $k$-strings with $k=\frac{1}{2}(N\pm1)$
must be identical. The same is true for $k=\frac{1}{2}(N\pm 2)$
if $N$ is even.

The $k$-string  tension 
cannot depend on particular representation
of the probe color source, but only on its $N$-ality.
Indeed, the particular Young tableau of the representation
plays no role, since all representations with the given $N$-ality
can be converted into each other through
emission of an appropriate number of soft gluons. For instance,
the adjoint representation has vanishing $N$-ality;  
the color source in the adjoint can be completely screened by gluons,
and the flux tube between the adjoint color sources should not
exist. The symmetric two-index representation
$Q^{\{ij\}}$ can be transformed into antisymmetric
$Q^{[ij]}$ plus a gluon, $Q^{\{ij\}}\to Q^{[ik]}+G_k^j$, and so on.

The above statement seemingly contradicts
the abundant lattice literature on
the  adjoint strings, measurements of distinct string tensions
for symmetric and anti-symmetric representations of one
and the same $N$-ality, and so on. Does it?

It turns out that at large $N$ some ``wrong" strings
are quasi-stable.  Being created, they must relax to
{\em bona fide} $k$-strings (i.e. those corresponding to 
sources with $k$ indices in fully antisymmetric representation),
but the relaxation time is exponentially large. 
Surprisingly, the question how fast they relax has never been solved previously.
I will present a sample estimate for the two-index symmetric representation
in Sect.~\ref{dpqss}. In this case to measure the {\em bona fide} $2$-string
tension one should deal with Wilson contours whose area is much larger than
exp$(\gamma N^2)$, where $\gamma$ is a numerical constant.

\section{Casimir vs. sine formula}
\renewcommand{\theequation}{\thesection.\arabic{equation}}
\setcounter{equation}{0}

If $\sigma_f$ is the tension of the fundamental string,
the Casimir formula reads
\beq
\sigma_k = \frac{C_R}{C_{\rm fund}} \,\sigma_{\rm f}
\eeq
where $C_R$ is the quadratic Casimir coefficient for representation $R$
defined as  $$T^aT^a = C_R\,I_R$$ (here $I$ is the unit matrix in the 
representation $R$,
while $T^a$'s stand for the SU$(N)$ generators in the same representation).
For antisymmetric $k$-index representation
\beq
\sigma_k = k\left( 1-\frac{k-1}{N-1}\right) \,\sigma_{\rm f}\,.
\label{cafo}
\eeq
The large-$N$ expansion of the Casimir formula is
\beq
\sigma_k = k\left( 1-\frac{k}{N} +O(N^{-2})\right) \,\sigma_{\rm f}\,.
\eeq
The expansion runs in even and odd powers of $1/N$.

\vspace{1mm}

Now, let us compare it with the sine law for the $k$-string tension
which reads
\beq
\sigma_k = \left(\sin\frac{\pi}{N} \right)^{-1} \sigma_{\rm f} \, \sin\left(\frac{\pi\,k}{N}\right)\,.
\eeq
At large $N$ the sine formula can be expanded as follows:
\beq
\sigma_k = k\left( 1-\frac{\pi^2}{6 N^2}\,(k^2-1) +O(N^{-4})\right) \,
\sigma_{\rm f}\,.
\eeq
The difference which immediately becomes obvious 
is that in the sine formula  the large-$N$ expansion runs
in even powers of $1/N$ while in the Casimir formula all powers of 
$1/N$ are involved. 

Let us ask ourselves what should one expect in Yang--Mills theory. 
Assume that we start from two distant fundamental strings, each  attached to
a (infinitely heavy)
probe quark at the top and a probe antiquark at the bottom, as in Fig.~\ref{st1}.
The distance between the probe quark and antiquark is $L\to\infty$.
the distance between two $Q$'s is $\ell$, and so is the
distance between two $\bar Q$'s.
No dynamical quarks are present in our theory. Then we 
let two $Q$'s adiabatically  approach each other, keeping the strings parallel, and
 eventually make the
distance $\ell$ between them less than $\Lambda^{-1}$. 

\begin{figure}
\centerline{\includegraphics[width=5in]{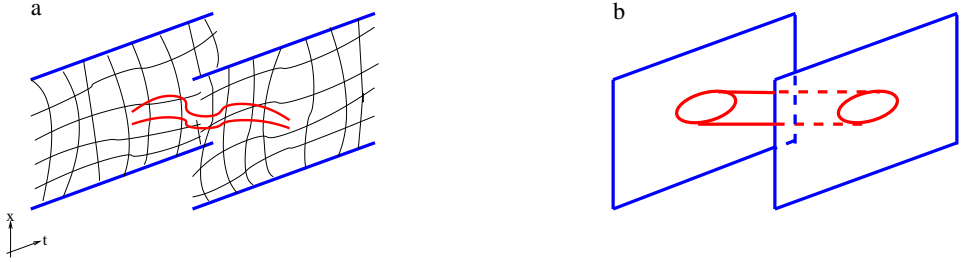}}
\caption{The interaction of two ``fundamental'' strings: (a)
Field-theory picture --- (two)gluon exchange; (b) String-theory picture ---
exchange of a closed string between two worldsheets.}
\label{interaction}
\end{figure}

At $N=\infty$ the energy of this configuration is $2\sigma_{\rm f} L+\,
L$-independent part which is irrelevant for our purposes. In this limit the two strings do not interact.
At finite $N$ interaction switches on. The spatial extent of interaction
is $\sim \Lambda^{-1}$. If we consider parts of strings in the central domain, far away from the endpoints, the interaction has no knowledge of quarks whatsoever 
(Fig.~\ref{interaction}a).
In fact, in supersymmetric gluodynamics one could
eliminate probe quarks at all,  putting the endpoints of the strings under consideration on two distant parallel domain walls. The large-$N$ structure of the interaction
is the same as in pure Yang--Mills. Hence, the $1/N$ expansion runs in even powers 
\cite{Armoni-two}.

One can arrive at the same conclusion from the  
 string theory side
too \cite{Armoni-two}, see Fig.~\ref{interaction}b.
Time axis is in the horizontal direction, we have two parallel fundamental string
worldsheets, and their interaction is due to the closed string exchange
(annulus diagram). It is quite obvious that this contribution
is proportional to $g_{\rm st}^2$ where $g_{\rm st} $ is the
string coupling constant. Since $g_{\rm st}\sim 1/N $
we see again that 
 the string interaction starts from $1/N^2$ and 
contains only even powers of $1/N$.

One can argue on
general grounds  that at distances $\gsim \Lambda^{-1}$
the fundamental strings attract each other \cite{Armoni-two}, while at
short distances there is a repulsion, so that 
composite $k$-strings do develop. 
The Casimir scaling as an exact formula
for the $k$-string tension  is excluded. 
The sine formula does have an appropriate  $1/N$ 
expansion, but this is certainly no proof. 

Needless to say, getting a clear-cut understanding of the status 
of the sine formula 
in large-$N$ QCD is highly desirable. 
Unfortunately, there was no dramatic progress in this issue since 2003. 
Still, two remarks are in order.

The first remark concerns the derivation of the $k$-string tension
via supergravity.  For the Maldacena--Nu\~{n}ez background
\cite{Maldacena:2000yy}  the sine formula was found to be exact \cite{Herzog:2001fq}.
At the same time,
for the Klebanov--Strassler background \cite{Klebanov:2000hb} 
the sine formula proved to be an excellent approximation, valid 
to a few percent accuracy,
but not exact. (Of course, the  even power  $1/N$ expansion applies
in both cases.)
Recently Butti et al. realized \cite{Butti} that the 
Klebanov--Strassler solution is a limiting case of an entire branch of solutions
which goes under the name of a ``baryonic branch." By varying the string 
coupling and some other parameters one can smoothly interpolate between the Klebanov--Strassler background
and the Maldacena--Nu\~{n}ez one. Correspondingly, the
$k$-string tension will change by a few percent, tending to the sine formula
in the limiting case of the Maldacena--Nu\~{n}ez background.

Another pertinent result I want to mention here is the one obtained in a toy
model, to be discussed in detail in Sect.~\ref{kkc}.

\section{Relaxation of quasi-stable strings}
\label{dpqss}
\renewcommand{\theequation}{\thesection.\arabic{equation}}
\setcounter{equation}{0}

In this section I will sketch an
estimate of the quasi-stable string decay rates. 
As was mentioned, for given $N$-ality $k$
the stable string is the one corresponding
to the  $k$-index representation with all indices totally antisymmetrized.
Assume that we ``prepared" a string corresponding to a different representation
with the same $N$-ality. How long does it take for the excited  string to relax?

To answer this question we consider 
strings of length $L$ in the Minkowski space-time.
Eventually we will take $L\to\infty$. The task is to 
 calculate the $N$ and representation
dependence of the
decay rates per unit time per unit length 
of the string. In some instances the answer was known long ago \cite{grin},
for instance, for the notorious ``adjoint string"
which should decay into a no-string state.
If the probe heavy sources are in the adjoint representation
they can be screened off   via  creation of a pair of gluons.
At the hadronic level, the string breaking
is equivalent to the statement that the operator $\bar Q^i_j \, Q^j_i$
produces a pair of (color-singlet) mesons of the type
$QG$. Here $Q^j_i$ is the field of the probe heavy quark
while $G$ stands for the gluon.
It is easy to see that the  probability
of the string breaking (per unit length per unit time) is
$\Lambda^2 /N^2$. This power-like suppression could be strong enough numerically
for such strings to show up on lattices as quasi-stable. Parametrically, it
is of the same order as the ``binding energy"
(which is also suppressed as $N^{-2}$). For this reason the tension of the ``adjoint string" is an ill-defined notion.

More challenging and less studied
are strings attached to color sources of the type
$Q^{i_1,i_2,..., i_k}$ or $Q_{i_1,i_2,..., i_k}$ in representations other than
fully antisymmetric. I will discuss just one example, 
 $k=2$. Other cases are considered in \cite{Armoni-two}.
In this case irreducible representations are of two types:
$Q^{\{ij\}}$ and $Q^{[ij]}$ (symmetric and antisymmetric).
The 2-string with the lowest tension corresponds to the
antisymmetric representation.  
Since the string interaction is $O(1/N^2)$
 the tension splitting between the symmetric quasi-stable
string (with a larger tension) and the antisymmetric one 
scales as $\Lambda^2\, N^{-2}$. 
Let us assume we start  from the  excited string, i.e.
consider  the Wilson loop for  $Q^{\{ij\}}$. 
I  will show that the symmetric string does not decay
into the antisymmetric one to any finite order in 
$1/N^2$. The decay rate is exponential.

Indeed, in order to convert the symmetric color representation into
the antisymmetric representation one has to produce a pair of gluons.
This takes energy of the order of $\Lambda$.
However, the string is not entirely broken,
rather it is restructured, with the tension splitting
$\sim \Lambda^2 N^{-2}$. To collect enough energy,
the gluon creation should take place not locally, but, rather
at the interval of the length $\sim \Lambda^{-1}\, N^2$.
This is  a typical tunneling process.

\begin{figure}
\centerline{\includegraphics[width=5in]{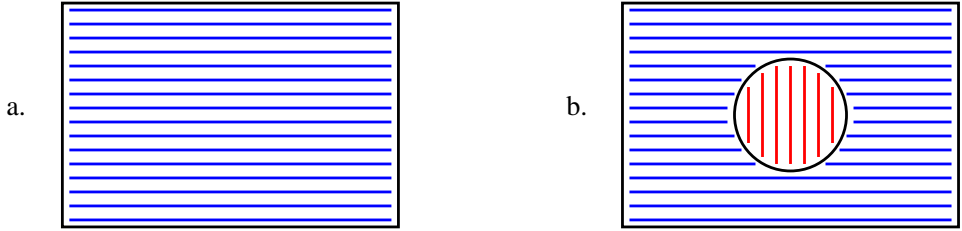}}
\caption{The world-sheet of a 2-string. (a). A purely symmetric string
(denoted by horizontal lines). (b). The decay into an antisymmetric
string (vertical lines) via an expanding bubble.}
\label{symm}
\end{figure}

The decay rate can be found quasiclassically.
The worldsheet of the symmetric string is shown in Fig. \ref{symm}a.
Its decay proceeds via a bubble creation, see Fig. \ref{symm}b.
The worldsheet of the symmetric string is two-dimensional 
``false vacuum,'' while inside the bubble
we have a ``true vacuum,'' i.e. the surface
spanned by the anti-symmetric string. The tension difference
--- in the false vacuum decay problem, the vacuum energy difference ---
is ${\cal E} \sim  \Lambda^2 N^{-2}$, while the energy $T$
of the bubble boundary
(per unit length) is $T\sim \Lambda$. This means that the thin wall approximation
is applicable, and the decay rate
$\Gamma$ (per unit length of the string per unit time)
is \cite{Kobzarev:1974cp,Voloshin:1995in}
\beq
\Gamma_{{\rm sym} \to {\rm antisym}} \sim \Lambda^2 \exp\left(-\frac{\pi\, T^2}{{\cal E}}\right)  
\sim \Lambda^2 \exp\left(-\gamma \, N^2 \right)\,,
\label{expdec} 
\eeq
where $\gamma$ is a positive constant of the order of unity.
Once the true vacuum domain is created through tunneling,
it will expand in   real time pushing the boundaries
(i.e. the positions of the gluelumps responsible for the conversion)
toward the string ends.

The above estimate relies on the assumption 
that the gluelump energy $\sim \Lambda$. 
This assumption seems natural, it is hard to imagine any other regime,
say, the gluelump mass scaling as $\Lambda/N$. It would be 
highly desirable  to get a better
understanding of this issue. Presumably, one should be able to write an
effective low-energy theory on the 2-string worldsheet which has two vacua:
one stable and one quasi-stable, whose energy density becomes degenerate with that of
the stable vacuum in the limit $N\to\infty$. A kink in this theory
would describe a gluelump. 
(See Sect.~\ref{naswc} for a related discussion.)
Although the {\sl domain wall} worldvolume theory 
of this type is known for quite some time \cite{Acharya},
I am aware of no attempts at deriving a QCD $k$-string worldvolume theory 
along these lines.

\section{If $L< L_{\rm crit} $
  excited strings may be stable}
\label{if}
\renewcommand{\theequation}{\thesection.\arabic{equation}}
\setcounter{equation}{0}

So far the vast majority of lattice simulations
yields string tensions which depend on the particular representation of
the probe source rather than on its $N$-ality, with the tension 
proportional to the Casimir coefficient of the representation 
at hand.\footnote{More on the current situation on lattices will be said in 
Sect.~\ref{ndol}.} The reason behind this situation may be due to
the fact that on lattices so far one deals with a mixture of
``wrong" excited strings with just a small admixture of the genuine 
ground-state $k$ string. The studies of the genuine  $k$
strings are hampered by the longevity of the
excited strings. As was argued in Sect.~\ref{dpqss},
excited strings live exponentially long. For, instance, the decay rate 
of the two-index
symmetric string into two-index antisymmetric was estimated as  
 $\Lambda^2 \exp\left(-\gamma \, N^2 \right)$.

The above estimate refers to the string of an infinite length.
Needless to say, when strings are treated on lattices, they have finite length.
If the distance between the probe sources is
less than some critical distance $L_{\rm crit}$
the excited string may turn out to be stable,
  protected against decay  into
the ground state by energy balance.

Indeed, the energy of the probe configuration
depicted in Fig.~\ref{st1}b consists of two parts:
the  string proper (this energy scales as $L$) and the end-point domains ---   
  ``bulges" ---
where the string couples to the color sources (this energy scales as $L^0$).
The energy of the bulges depends, generally speaking, on the particular
representation of the color sources, as well as on the structure of the
string proper, i.e. symmetric versus antisymmetric. The bulge connecting
the symmetric source with the antisymmetric string
is expected to be heavier than that connecting
the symmetric source with the symmetric (quasi-stable) string since it absorbs
a gluelump. The crucial question is the bulge energy difference
in these two cases.  If it is $\sim \Lambda$, we need the string
to be longer than $\Lambda^{-1} N^2$ for the energy gain
due to the string decay  to overcome
the energy loss in the end-point domains. Since the bulge is ``locally" coupled to the
color source it is conceivable that the bulge energy difference scales as
$\sim \Lambda /N$. Then the critical string length is 
$  L_{\rm crit} \gsim \Lambda^{-1} N$.

Thus,  ``finite-length excited  strings" may not be able to
relax if the lattice size is smaller than  $L_{\rm crit}$. In this case  of ``short" strings
it will be
impossible, even  in principle, to measure the genuine $k$-string tension
no matter how long is the time extension. This circumstance was noted by Gliozzi
\cite{Gli}. I merely adapt his argument within the framework of the large-$N$ analysis.

It is important that 
$  L_{\rm crit}\gg \Lambda^{-1}$. Otherwise framing the discussion in terms of
strings would  not be appropriate at all.

\section{Lattices: what's happening?}
\label{ndol}
\renewcommand{\theequation}{\thesection.\arabic{equation}}
\setcounter{equation}{0}

Numerical study of $k$ strings   is being pursued,
for quite some time, by a number of lattice practitioners, of which I will mention here
two groups: one in Italy and one in UK (see e.g. \cite{DD,LT}).
By and large, the first group  finds agreement with the sine law, while the second one obtains results lying between 
the sine law and the Casimir formula, with somewhat larger errors, which cover both
possibilities. A report on the  recent progress of the second group can be found in 
\cite{LTW}.
Lucini et al. obtain
string tensions which are somewhat  smaller than those measured 
by the first group,  for the same values of lattice parameters,
which  indicates, generally speaking,  that  separation of  the ground-state 
string is better in this case. As usual, a potentially dangerous 
 source of systematic errors is the finite-volume correction:
both groups rely on the leading bosonic string correction
$\sigma_{\rm eff}(L) = \sigma - \pi\, L^{-2}/3 $
for extrapolation of their data obtained on a size $L$  lattice to $L\to\infty$.   
It is not quite clear  whether subleading corrections could
appreciably distort the ratio $\sigma_k/\sigma_{\rm f}$.

In any case, the real question is about the $N\to\infty$ limit, not about $N=4,\,\, 6,\,\,8$.
A clean way to address this question is to consider, say, $\sigma_2/\sigma_{\rm f}$
as a function of $N$, and check whether  this ratio receives $1/N$ or $1/N^2$ leading
corrections. The first analysis of this type was performed in \cite{LTW}, see
Eq.~(42) and Fig.~15 in this paper. The errors are way too large
to provide us with a conclusive answer. To get a  conclusive answer large values of $N$ and high accuracy are needed. Experts say they 
do not expect this issue to be resolved any time  soon through this procedure.

At the same time, it  is gratifying to note that a number of lattice practitioners
stimulated by the pioneering work \cite{PdP} (where a new method for studying the
adjoint string breaking was suggested)
became interested in developing new approaches to 
quasi-stable strings and string decays. Shortly after
the original publication \cite{PdP} this
issue was revisited in \cite{DDP} for the
symmetric two-index representation  of SU(3), which allowed the authors to explain
earlier  results that had been reported in \cite{DDPR,DD}. A fresh theoretical analysis along these lines is presented in \cite{Gli}.

Concluding this section let me mention, as a side remark,
a  landmark achievement marginally related to my current topic: a detailed description
of the fundamental string breaking through quark-antiquark pair creation in full QCD was recently  obtained in  \cite{Bali}.

\section{Non-Abelian strings at weak coupling}
\label{naswc}
\renewcommand{\theequation}{\thesection.\arabic{equation}}
\setcounter{equation}{0}

In this section I describe seemingly the simplest {\em weakly coupled}
model in which the Meissner effect does take place and
leads to formation of {\em non-Abelian} chromomagnetic flux tubes.
The model supports non-Abelian confined magnetic monopoles.
In the dual description the magnetic flux tubes are prototypes
of QCD strings. Dualizing the confined magnetic monopoles
we get gluelumps (string-attached gluons)
which convert a ``QCD string" in the excited state
to that in the ground state. The decay rate of the excited string
to its ground state is suppressed exponentially in $N$, much in the same way as in QCD,
see Sect. \ref{dpqss}.

The model we will discuss is in a sense minimal. Due to its weak coupling it is fully
controllable.
One can think of it as of a reference
model. 
It is non-supersymmetric (but can be readily supersymmetrized
if necessary). In Part II I will deal with a  (slightly broken)
supersymmetric version of the same model.

\subsection{The bulk structure of the toy four-dimensional Yang--Mills  model \cite{recent,Hanany,Auzzi,ShifmanYung,GSY,Tong,HananyTong,Markov}}

The gauge group of the model
is SU($N)\times$U(1). Besides SU($N$) and U(1)
gauge bosons   
the model contains $N$ scalar fields charged with respect to
U(1) which form $N$ fundamental representations of SU($N$).
It is convenient to write these fields in the form of 
$N\times N$ matrix $\Phi =\{\varphi^{kA}\}$
where $k$ is the SU($N$) gauge index while $A$ is the flavor
index, 
\beq
\Phi =\left(
\begin{array}{cccc}
\varphi^{11} & \varphi^{12}& ... & \varphi^{1N}\\[2mm]
\varphi^{21} & \varphi^{22}& ... & \varphi^{2N}\\[2mm]
...&...&...&...\\[2mm]
\varphi^{N1} & \varphi^{N2}& ... & \varphi^{NN}
\end{array}
\right)\,.
\label{phima}
\eeq
The action of the model has the form
\beqn
S &=& \int {\rm d}^4x\left\{\frac1{4g_2^2}
\left(F^{a}_{\mu\nu}\right)^{2}
+ \frac1{4g_1^2}\left(F_{\mu\nu}\right)^{2}
 \right.
 \nonumber\\[3mm]
&+&\left.
 {\rm Tr}\, (\nabla_\mu \Phi)^\dagger \,(\nabla^\mu \Phi )
+\frac{g^2_2}{2}\left[{\rm Tr}\,
\left(\Phi^\dagger T^a \Phi\right)\right]^2
 +
 \frac{g^2_1}{8}\left[ {\rm Tr}\,
\left( \Phi^\dagger \Phi \right)- N\xi \right]^2  \right\},
\label{redqed}
\eeqn
where $T^a$ stands for the generator of the gauge SU($N$),
\beq
\nabla_\mu \, \Phi \equiv  \left( \partial_\mu -\frac{i}{\sqrt{ 2N}}\; A_{\mu}
-i A^{a}_{\mu}\, T^a\right)\Phi\, .
\label{dcde}
\eeq
The global flavor SU($N$) transformations then act on $\Phi$ from the 
right. The action (\ref{redqed}) in 
fact represents a truncated bosonic sector of the ${\cal N}=2$
supersymmetric  model. The last 
term in the second line
forces $\Phi$ to develop a vacuum expectation value (VEV) while the 
last but one term
forces the VEV to be diagonal,
\beq
\Phi_{\rm vac} = \sqrt\xi\,{\rm diag}\, \{1,1,...,1\}\,.
\label{diagphi}
\eeq
To ensure weak coupling one must assume $\xi\gg\Lambda^2$.

The  vacuum field (\ref{diagphi}) results in  the spontaneous
breaking of both gauge and flavor SU($N$)'s.
A diagonal global SU($N$) survives, however,
namely
\beq
{\rm U}(N)_{\rm gauge}\times {\rm SU}(N)_{\rm flavor}
\to {\rm SU}(N)_{\rm diag}\,.
\eeq
Thus, color-flavor locking takes place in the vacuum.

This model has a string solution, which I will briefly review now.
Since it includes a spontaneously broken gauge U(1),
the model supports
conventional ANO strings  
in which one can discard the SU($N$)$_{\rm gauge}$ part 
of the action.
 These are not the strings we are interested in.
At first sight the triviality of the homotopy group, $\pi_1 ({\rm SU}(N)) =0$, 
implies that there are no other topologically stable strings.
This impression is false. One can
combine the $Z_N$ center of SU($N$) with the elements $\exp (2\pi i k/N)\in$U(1) 
to get a topologically stable string solution
possessing both windings, in SU($N$) and U(1), namely,
\beq
\pi_1 \left({\rm SU}(N)\times {\rm U}(1)/ Z_N
\right)\neq 0\,.
\eeq
It is easy to see that this nontrivial topology amounts to winding
of just one element of $\Phi_{\rm vac}$, say, $\varphi^{11}$, or
$\varphi^{22}$, etc, for instance
\beq
\Phi_{\rm string} = \sqrt{\xi}\,{\rm diag} ( 1,1, ... ,1,e^{i\alpha (x) })\,,
\quad x\to\infty \,,
\label{ansa}
\eeq
($\alpha$ is the angle of
the coordinate  $\vec{x}_\perp$ in the perpendicular plane.)
Such strings are referred to as elementary $Z_N$ strings.
They are progenitors of the non-Abelian strings.
Their tension is $1/N$-th of that of the ANO string, see Eqs.~(\ref{ten}) and 
(\ref{tenANO}).
The ANO string can be viewed as a bound state of 
$N$ elementary strings at a certain point in the moduli space.

\subsection{Elementary $Z_N$ strings}

At finite distances from the flux tube center the $Z_N$ string solution 
 can be written as
 \cite{Auzzi}
\beqn
\Phi &=&
\left(
\begin{array}{cccc}
\phi(r) & 0& ... & 0\\[2mm]
...&...&...&...\\[2mm]
0& ... & \phi(r)&  0\\[2mm]
0 & 0& ... & e^{i\alpha}\phi_{N}(r)
\end{array}
\right) ,
\nonumber\\[5mm]
A^{{\rm SU}(N)}_i &=&
\frac1N\left(
\begin{array}{cccc}
1 & ... & 0 & 0\\[2mm]
...&...&...&...\\[2mm]
0&  ... & 1 & 0\\[2mm]
0 & 0& ... & -(N-1)
\end{array}
\right)\, \left( \partial_i \alpha \right) \left[ -1+f_{NA}(r)\right] ,
\nonumber\\[5mm]
A^{{\rm U}(1)}_i &=& \frac{1}{N}\, 
\left( \partial_i \alpha \right)\left[1-f(r)\right] ,\qquad A^{{\rm U}(1)}_0=
A^{{\rm SU}(N)}_0 =0\,,
\label{znstr}
\eeqn
where $i=1,2$ labels coordinates in the plane orthogonal to the string
axis and $r$ and $\alpha$ are the polar coordinates in this plane. The profile
functions $\phi(r)$ and  $\phi_N(r)$ determine the profiles of the scalar fields,
while $f_{NA}(r)$ and $f(r)$ determine the SU($N$) and U(1) fields of the 
string solutions, respectively. These functions satisfy the following 
rather obvious boundary conditions:
\beqn
&& \phi_{N}(0)=0,
\nonumber\\[2mm]
&& f_{NA}(0)=1,\;\;\;f(0)=1\,,
\label{bc0}
\eeqn
at $r=0$, and 
\beqn
&& \phi_{N}(\infty)=\sqrt{\xi},\;\;\;\phi(\infty)=\sqrt{\xi}\,,
\nonumber\\[2mm]
&& f_{NA}(\infty)=0,\;\;\;\; \; f(\infty) = 0
\label{bcinfty}
\eeqn
at $r=\infty$.
Because our model is, in fact, a bosonic reduction
 of the ${\cal N}=2$ supersymmetric theory,
these profile functions satisfy the first-order differential equations, 
\beqn
&&
r\frac{d}{{d}r}\,\phi_N (r)- \frac1N\left( f(r)
+  (N-1)f_{NA}(r) \right)\phi_N (r) = 0\, ,
\nonumber\\[4mm]
&&
r\frac{d}{{ d}r}\,\phi (r)- \frac1N\left(f(r)
-  f_{NA}(r)\right)\phi (r) = 0\, ,
\nonumber\\[4mm]
&&
-\frac1r\,\frac{ d}{{ d}r} f(r)+\frac{g^2_1 N}{4}\,
\left[(N-1)\phi(r)^2 +\phi_N(r)^2-N\xi\right] = 0\, ,
\nonumber\\[4mm]
&&
-\frac1r\,\frac{d}{{ d}r} f_{NA}(r)+\frac{g^2_2}{2}\,
\left[\phi_N(r)^2 -\phi_2(r)^2\right]  = 0\, .
\label{foe}
\eeqn
These equations  can be solved numerically. Clearly, the solutions
to the first-order equations automatically satisfy the second-order equations 
of motion.\footnote{Quantum corrections destroy  fine-tuning of the coupling constants in
(\ref{redqed}). If one is interested in  calculation of the quantum-corrected 
profile functions one has to solve the second-order equations of motion
instead of (\ref{foe}).}

The tension of the elementary $Z_N$ string is 
\beq
T_1=2\pi\,\xi\, ,
\label{ten}
\eeq
to be compared with 
the tension of 
the ANO
string, 
\beq
T_{\rm ANO}=2\pi\,N\,\xi\,.
\label{tenANO}
\eeq

\subsection{Why non-Abelian?}

In which sense the elementary $Z_N$ strings are 
progenitors of the {\em bona fide} non-Abelian strings?
At the classical level they are all degenerate and can be continuously deformed to
one another. Indeed, 
besides trivial translational moduli, they have SU$(N)$ ``orientational"  moduli corresponding to spontaneous
breaking of a non-Abelian symmetry. Indeed, while the ``flat"
vacuum is SU($N$)$_{\rm diag}$ symmetric, the solution (\ref{znstr})
breaks this symmetry 
down to U(1)$\times$SU$(N-1)$ (at $N>2$).
This means that the world-sheet (two-dimensional) theory of 
the elementary string moduli
is the SU($N$)/(U(1)$\times$ SU($N-1$)) sigma model.
This is also known as $CP(N-1)$ model.

To obtain the non-Abelian string solution from the $Z_N$ string 
(\ref{znstr}) we apply the diagonal color-flavor rotation  preserving
the vacuum (\ref{diagphi}). To this end
it is convenient to pass to the singular gauge where the scalar fields have
no winding at infinity, while the string flux comes from the vicinity of  
the origin. In this gauge we have
\beqn
\Phi &=&
U\left(
\begin{array}{cccc}
\phi(r) & 0& ... & 0\\[2mm]
...&...&...&...\\[2mm]
0& ... & \phi(r)&  0\\[2mm]
0 & 0& ... & \phi_{N}(r)
\end{array}
\right)U^{-1}\, ,
\nonumber\\[5mm]
A^{{\rm SU}(N)}_i &=&
\frac{1}{N} \,U\left(
\begin{array}{cccc}
1 & ... & 0 & 0\\[2mm]
...&...&...&...\\[2mm]
0&  ... & 1 & 0\\[2mm]
0 & 0& ... & -(N-1)
\end{array}
\right)U^{-1}\, \left( \partial_i \alpha\right)  f_{NA}(r)\, ,
\nonumber\\[5mm]
A^{{\rm U}(1)}_i &=& -\frac{1}{N}\, 
\left( \partial_i \alpha\right)   f(r)\, , \qquad A^{{\rm U}(1)}_0=
A^{{\rm SU}(N)}_0=0\,,
\label{nastr}
\eeqn
where $U$ is a matrix $\in {\rm SU}(N)$. This matrix parametrizes 
orientational zero modes of the string associated with flux rotation  
in  SU($N$). The presence of these modes makes the string genuinely
non-Abelian. Since the diagonal color-flavor symmetry is not
broken by the vacuum expectation values 
(VEV's) of the scalar fields 
in the bulk (color-flavor locking)
it is physical and has nothing to do
with the gauge rotations eaten by the Higgs mechanism. The orientational moduli
encoded in the matrix $U$ are {\it not} gauge artifacts. The orientational
zero modes were first 
observed in \cite{Hanany, Auzzi}.

Thus, at the classical level there is a continuous manifold of
non-Abelian strings. The degeneracy is lifted at the
quantum level.

\subsection{Structure of the string worldsheet theory}

As is clear from the string solution (\ref{nastr}),   not each element of
the matrix $U$ will give rise to a modulus. The SU($N-1) \times$U(1) subgroup 
remains unbroken 
by the string solution under consideration; therefore,  as  was already mentioned,
the moduli space is 
\beq
\frac{{\rm SU}(N)}{{\rm SU}(N-1)\times {\rm U}(1)}\sim CP(N-1)\,.
\label{modulispace}
\eeq
Keeping this in mind we parametrize the matrices entering Eq.~(\ref{nastr})
as follows:
\beq
\frac1N\left\{
U\left(
\begin{array}{cccc}
1 & ... & 0 & 0\\[2mm]
...&...&...&...\\[2mm]
0&  ... & 1 & 0\\[2mm]
0 & 0& ... & -(N-1)
\end{array}
\right)U^{-1}
\right\}^l_p=-n^l n_p^* +\frac1N \delta^l_p\,\, ,
\label{n}
\eeq
where $n^l$ is a complex vector
 in the fundamental representation of SU($N$), and
\beq
 n^*_l n^l =1\,,
 \label{dddd}
\eeq
($l,p=1, ..., N$ are color indices).
One U(1) phase is gauged in the effective
sigma model. This gives the correct number of degrees of freedom,
namely, $2(N-1)$. 

Using this parametrization it is not difficult to obtain \cite{GSY}
an effective (1+1)-dimensional action for the moduli fields $n^i$,
\beq
S^{(1+1)}= 2 \beta\,   \int d t\, dz \,  \left\{(\partial_{\alpha}\, n^*_l
\partial_{\alpha}\, n^l) + (n^*_l\, \partial_{\alpha}\, n^l)^2\right\}\,,
\label{o3}
\eeq
where the coupling constant $\beta$ is given by 
\beq
\beta=\frac{2\pi}{g^2_2}\,.
\label{beta}
\eeq
Equations (\ref{o3}) and (\ref{dddd})
present one of many various parametrizations of CP($N-1$) model which are in use in the literature. It is well known that the continuous degeneracy
of the classical vacuum manifold of CP($N-1$) model is lifted at the quantum 
level (for a pedagogical discussion see e.g. \cite{5}). The quantum vacuum is unique.
This means that, unlike $Z_N$ strings, our non-Abelian four-dimensional 1-string
is unique. This is good. Other aspects are not so good, however.

In two-dimensional CP($N-1$) model there is a large number (of the order of $N$)
of quasistable ``vacuum states" of the
type depicted in Fig.~\ref{dva}.  The quasistable vacua of CP($N-1$) model decay into the genuine one with probability $\sim \Lambda^2 \exp\left(-\gamma \, N \right)$, see 
Sect.~\ref{gmv}. 
In terms of four-dimensional strings this means the
existence of a large number of ``excited strings" 
which have nothing to do with the excited $k$-strings
since the quasistable vacua of CP($N-1$) model appear for four-dimensional 1-strings,
and, moreover, the energy-density spacing is $\Lambda^2/N$.
Thus, the parallel with QCD is far from being perfect. It is fair to say that
we made just the first little step in a long journey which may or may not
lead to adequate modeling of QCD strings at weak coupling.

Note that at
large $N$ the worldsheet CP($N-1$) model is solvable. Supersymmetrization of the model
which will be needed in Part II can be carried out
following the program  of \cite{ShifmanYung}.

\vspace{1cm}

{\huge Part II: Original}

\section{
$k$-kink confinement in two-dimensional\\
\boldmath{$CP(N-1)$} models: hints for $k$-string tension?} 
\label{kkc}
\renewcommand{\theequation}{\thesection.\arabic{equation}}
\setcounter{equation}{0}

It is known from the early 1970's that
four-dimensional Yang-Mills theory and two-dimensional O(3)
sigma model exhibit remarkable parallels. 
Just like  Yang--Mills theory,  
O(3) sigma model has asymptotic freedom 
\cite{1} and instantons \cite{2}. Moreover, $CP(N-1)$
sigma models introduced in Ref.~\cite{3,6}
present a parallel to large-$N$ Yang--Mills (YM) theories. 
The parameter $N$ of $CP(N-1)$ plays the same 
role as the number of colors in QCD,
for instance, $CP(1)$ (equivalent to $O(3)$) is 
analogous to SU(2) Yang-Mills,
$CP(2)$ to SU(3) and so on. The large-$N$ expansion in the 
sigma models was analyzed in \cite{4,5},
with the conclusion that its structure is very similar to that
in YM theories.   Finally, 2D supersymmetric $CP(N-1)$
sigma model constructed in \cite{6} is an
excellent  toy model
for 4D supergluodynamics (see e.g. the review paper
\cite{NSVZsigma}).

The similarity of the strong-coupling dynamics
of four-dimensional Yang--Mills theories and
two-dimensional $CP(N-1)$
sigma models is not accidental. As was discussed in Sect.~\ref{naswc}
some Yang--Mills models support  
non-Abelian strings 
with moduli whose worldsheet interaction is
described by $CP(N-1)$. In other words, these YM models are microscopic 
theories of phenomena 
whose macroscopic description is given by 
two-dimensional $CP(N-1)$ models. The Meissner mechanism 
of confinement translates into quite certain statements 
concerning kinks in two-dimensional $CP(N-1)$.
This explains  observations of a number of parallel
phenomena in the two classes of theories, 2D and 4D.
I want  to further exploit this parallelism  to  pose a 
question in $CP(N-1)$ model which seemingly escaped theorists' attention
so far.

The issue I want to address is  the  tension
 of the
 configuration with a $k$-kink and $k$-antikink, $k$-kink tension for short.
In broken
supersymmetric $CP(N-1)$ kinks are confined, much in the same way as quarks in QCD.
I will show that the sine formula
for the $k$-kink tension emerges  in a 2D $CP(N-1)$ model.
As usual with toy models, results obtained in toy
models, strictly speaking,  prove nothing with regards to the real 
thing. However, with luck, they can provide insights
which might be helpful in addressing the original theory.
Moreover, although on the one hand toy models are poorer than the original
theory (e.g.  in the case at hand it is two-dimensional), on the other
hand, they may contain free parameters which
allow us to fine-tune them
so that a variety of dynamical regimes
becomes accessible. In the $CP(N-1)$ model we will deal with,
several distinct regimes can be realized; one of them is somewhat
similar to that of QCD. The interplay between various regimes, including
the QCD-like regime, is quite fascinating. Finally, let me mention that
$CP(N-1)$ models have a value of their own, unrelated to QCD parallels.
To the best of my knowledge, the aspect of the $CP(N-1)$ models I will consider here
has never been addressed previously.

\subsection{CP($N-1$) model: nonlinear formulation}

I start from a brief introduction to 2D supersymmetric
${\rm  CP}(N-1) $ models,
with emphasis on features that will be exploited 
in this work.

For any  K\"{a}hler target space endowed with the metric
$G_{\bar j\, i}$ the Lagrangian of the  ${\cal N}=2$ model is
\beqn
{\cal L}_0 =
G_{\bar j\, i}\, \partial^\mu\bar\phi^{\,\bar j}\, \partial_\mu\phi^i
+\frac{i}{2} G_{\bar j\, i}\,
\Psi^{\dagger \bar j} \!\stackrel{\leftrightarrow}{\not\! \! D}\Psi^{i}
+R_{\bar j i k\bar{l}}\,
\Psi_L^{\dagger\bar j}\Psi_L^i\Psi_R^k \Psi_R^{\dagger\bar l}
\,,
\label{one}
\eeqn
where  $D_\mu$ is the covariant derivative,
\beq
 D_\mu\,\Psi^i = \partial_\mu\,\Psi^i +
\partial_\mu{\phi}^{ k}\, \Gamma_{ kl}^i\,\Psi^l
\,,
\label{two}
\eeq
$\Gamma$'s stand for the
Christoffel symbols,  $R_{\bar{j} i k\bar{l}}$
is the curvature tensor, while $\Psi$ denotes a two-component spinor,
\beq
\Psi =\left(\begin{array}{cc}
\Psi_R \\
\Psi_L
\end{array}
\right),
\eeq
so that the kinetic term of the fermions can be identically rewritten as
\beq
\frac{i}{2} G_{i\bar j}\,
\Psi^{\dagger\bar j} \!\stackrel{\leftrightarrow}{\not\! \! D}\Psi^{i}
=
\frac{i}{2}
G_{i\bar j}\left\{\Psi_L^{\dagger\bar j}
  \!\stackrel{\leftrightarrow}{\not\! \! D}_R\Psi_L^i
+ \Psi_R^{\dagger\bar j}  \!\stackrel{\leftrightarrow}{\not\! \! D}_L\Psi_R^i
\right\}\,.
\label{three}
\eeq
The right and left derivatives $\partial_{R,L}$ are
\beq
\partial_R =
\partial_0- \partial_1\,,\qquad \partial_L = \partial_0 + \partial_1\,.
\label{rlder}
\eeq
The metric is obtained from the K\"{a}hler potential
by differentiation,
\beq
G_{i\bar j} =\frac{\partial }{\partial \phi^i }\, \frac{\partial }{\partial  \phi^{\dagger\, \bar j} }\,
K (\phi ,\phi^\dagger )\,.
\eeq
The K\"ahler potential giving rise to the CP($N-1$) model
with the metric in the Fubini-Studi form
is
\beq
K_{{\rm CP}(N-1)} = \frac{2}{g^2}\ln\left( 1 + \sum_{i=1}^{N-1} |\phi^i|^2
\right)\,.
\eeq
Here $g^2$ is the dimensionless coupling constant of the
model. 
The CP($N-1$) metric can be written as
\beq
G_{i\bar j} = \frac{2}{g^2}\, \delta_{i\bar j}\, \frac{1}{\left( 1 + \sum_{i=1}^{N-1} |\phi^i|^2\right)} + \left( \phi^{\dagger\, \bar i}\, \phi^j \right)  \mbox {-term}\,.
\eeq
In what follows it will be useful to use the fact \cite{KN} that
in generic ${\cal N}=2$ compact homogeneous symmetric
K\"{a}hler sigma models  the Ricci tensor  $ R_{\bar j\, i}$ is proportional to the metric. In the case of the CP($N-1$) model
\beq
 R_{\bar j\, i} = N\,\frac{g^2}{2}    G_{i\bar j}  \,.
\eeq
(In the general case the coefficient $N$ is replaced by
$b$, the first --- and the only --- coefficient in the Gell-Mann--Low function.)

In addition, one can introduce a
$\theta$ term which has the following  form
\beq
{\cal L}_\theta =\frac{i\theta}{2\pi}\, \frac{g^2}{2}    G_{i\bar j} \,\,
 \varepsilon^{\mu\nu}
\partial_\mu \phi^{\dagger\, \bar j}\partial_\nu\phi^i \,.
\label{four}
\eeq
If supersymmetry is unbroken, one can always rotate the
$\theta$ term away using the anomaly in the axial current;
the $\theta$ angle is unobservable.
For a detailed review see \cite{CP(N-1)}.

As we will discuss shortly, this model has $N$ discrete degenerate vacua.
The excitation spectrum of the model consists
of kinks interpolating between these vacua.

We will also need to introduce a small
supersymmetry (SUSY) breaking. To this end
a SUSY-breaking 
mass term for fermions  will be added
to the Lagrangian (\ref{one}) and used  in due time.
The SUSY-breaking 
mass term can be parametrized as follows:
\beq
{\cal L}_m = 
-i\, m  G_{\bar j\, i}\,\Psi^{\dagger\bar j}_L \Psi^{i}_R 
+\mbox{h.c.} = -i\, m \, 
\frac{2}{Ng^2}\left(   R_{\bar j\, i}\,
\Psi^{\dagger \bar j}_L \Psi^{i}_R\right)
 +\mbox{h.c.}\,,
 \label{susybr}
\eeq
where $m$ is a complex parameter,
\beq
m = \mu e^{i\alpha}\,,\qquad \mu > 0\,,
\eeq
and $Ng^2$ is the 't Hooft coupling.
The combination $\mu/g^2$ is scale independent,
and so is $R\bar\Psi\Psi$.
If $\mu\neq 0$ the $\theta$ term becomes observable.
In fact, it is sufficient to keep
only one of two parameters, $\theta \neq 0$ or $\alpha \neq 0$,
since by a chiral rotation of the $\Psi$ fields one can always make
$m$ real at a price of an appropriate  shift of
$\theta$.
For technical reasons it is
slightly more convenient to keep $\alpha$ at zero, 
while considering $\theta $, the vacuum angle,  as a free parameter.
From now on we will assume $m$ to be real and positive.

\subsection{Linear gauged  formulation}

Some physical aspects of the $CP(N-1)$ model
become more transparent if one uses $N$ constrained
fields $n^i$ (this is the formulation discussed by Witten \cite{5}).
In this language the Lagrangian is built from an $N$-component
complex field $n^i$ subject to constraint
\beq
n_i^*\, n^i =1\,,
\label{lambdaco}
\eeq
(see e.g. \cite{NSVZsigma}),
plus  an $N$-component Dirac Fermi field
$\psi^i$ subject to the constraint
\beq
n_i^*\, \psi^i =0\,.
\label{chico}
\eeq
The Lagrangian has the form
\beqn
{\cal L} &=& \frac{2}{g^2}\, \left[
\left(\partial_\mu - i A_\mu\right) n^*_i
\left(\partial_\mu + i A_\mu\right) n^i
+\bar\psi_i (i\not\! \partial - \not\!\! A )\psi^i
-\frac{1}{2}\left(\sigma^2+\pi^2\right) \right.
\nonumber\\[2mm]
&-&\left. 
\frac{1}{\sqrt 2}\,\,\bar\psi \left(\sigma +i\pi\gamma_5
\right)\psi -\lambda \left( n^*_i n^i-1\right) +\bar\chi
 n^*_i \psi^i + \bar\psi_i n^i \chi\right] ,
\eeqn
where $\chi$ and $\bar\chi$ are Lagrange multiplier fields
that enforce the constraint
(\ref{chico}), $\lambda$ is the Lagrange multiplier
enforcing (\ref{lambdaco}) while $A_\mu$, $\sigma$ and $\pi$
are auxiliary fields which could be eliminated through equations of motion.
At the quantum level the above constraints will be gone.

The fields $n_i$ and $\psi^i$ are $N$-plets.
In the formulation (\ref{one}) they are 
represented by ``basic" kinks interpolating between the adjacent
vacua of the model, the so-called 1-kinks see below. For this reason, 
from now on I will refer to the $n^i$ quanta as to 
kinks.  Loosely speaking,  kinks $\leftrightarrow$
QCD quarks. The $n^i$ kink $N$-plet
corresponds to the $N$-plet of fundamental quarks.

In this representation the $\theta$ term 
can be written as
\beq
{\cal L}_\theta = \frac{\theta}{2\pi }\, \varepsilon_{\mu\nu} \partial^\mu A^\nu
=\frac{\theta}{2\pi }\, \varepsilon_{\mu\nu} \partial^\mu \left(
n_i^*\partial^\nu n^i
\right)\,.
\eeq

Witten showed, by exploiting the $1/N$ expansion to the leading order, 
that the kink mass is dynamically generated,
\beq
M_{\rm kink}^2 
=\Lambda^2 \equiv M_0^2 \exp\left(-\frac{8\pi}{Ng^2}\right).
\label{scp}
\eeq
Here $M_0$ is the ultraviolet cut off and $g^2$ is the bare
coupling constant. The combination $Ng^2$ is nothing but
the 't Hooft constant that does not scale with $N$. 
As a result, $M_{\rm kink}$ scales as $N^0$  at large $N$.
This result will be confirmed below in a different way.

In the non-supersymmetric version of the $CP(N-1)$ model
Witten found that kinks are subject to confinement,
the confining potential grows linearly with distance, with
the tension suppressed by $1/N$,
\beq
\sigma \sim \frac{M_{\rm kink}^2}{N}
\,.
\eeq
In other words,  confinement considered by Witten becomes
exceedingly weak at large $N$. The kink-antikink system can be described
by a non-relativistic Schr\"{o}dinger-like 
equation. This regime does not resemble
QCD where the string tension does not vanish at $N\to\infty$.
Since we are after emulating QCD we will have to look for
another regime. In QCD-like regime the string tension should scale
as $N^0$.

According to Witten's $1/N$ analysis which has been
repeatedly mentioned above \cite{5}, 
in the supersymmetric version of the $CP(N-1)$ model, the tension  vanishes; there is no kink confinement. Thus, it is clear that supersymmetry must be broken.

\subsection{What did people learn after Witten?}

\begin{figure}
 \centerline{\includegraphics[width=2in]{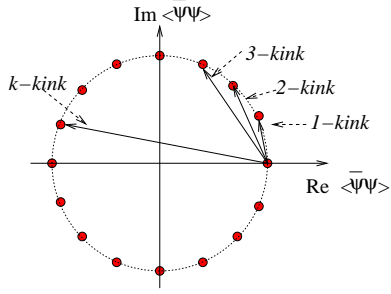}}
\caption{
The fermion condensate $ \left\langle i\, R_{\bar j\, i}\,
\bar\Psi^{\bar j}_L \Psi^{i}_R
\right\rangle$
 is the 
order parameter labeling distinct vacua in $CP(N-1)$ model
(the example shown corresponds to $N=16$).}
\label{ndsv}
\end{figure}

The development which is most relevant
for what follows is the determination of
the fermion condensate
\beq
\left\langle i\, \sum_{j,k}
 R_{\bar j\, k}\,
\Psi^{\dagger \bar j}_L \Psi^{k}_R
\right\rangle \neq 0\,,\qquad j,k = 1,2, ... , N-1\,,
\label{conde}
\eeq
which is the order parameter exhibiting the
spontaneous breaking of $Z_{2N}$ symmetry of the
$CP(N-1)$ model down to $Z_2$. The $Z_{2N}$
is a discrete remnant of the anomalous axial U(1).
The condensate (\ref{conde}) scales as $N^1$.
It can be exactly calculated, see e.g.
\cite{NSVZsigma}, and takes $N$ distinct values, much
in the same way
as the gluino 
condensate $\langle {\rm Tr}\lambda\lambda\rangle$
in SU($N$) SYM four-dimensional theory \cite{SV}. 
Namely,
\beq
\left\langle i\, \sum_{j,k}
 R_{\bar j\, i}\,
\Psi^{\dagger \bar j}_L \Psi^{k}_R
\right\rangle
= - N\Lambda \exp\left(\frac{2\pi\, i\, k}{N} +i\frac{\theta}{N}
\right)\,,\qquad k=0,1, ... N-1\,,
\label{nvalue}
\eeq
where $\Lambda$ is the scale parameter, see (\ref{scp}).
Equation (\ref{nvalue})
refers to arbitrary $\theta$. The dependence on $\theta /N$,
in conjunction with the physical $2\pi$ periodicity in $\theta $,
prompts us that in the model at hand there are $N$ vacua.
In fact, the fermion condensate (\ref{nvalue}) is the order parameter; it
labels $N$ discrete supersymmetric vacua of the
$CP(N-1)$ model, see Fig. \ref{ndsv}. Needless to say, in all  
 vacua the energy density  vanishes as a consequence of supersymmetry.

The second relevant development is the fact \cite{Loshif} (see also \cite{5,SVZ})
that the above operator appears as an anomalous   central charge
in the superalgebra,
\beqn
\{  Q_R^\dagger \,,\,  Q_L \} &=&
\frac{i}{2 \pi}  \int dx\, \partial_x \left( R_{i\bar j}\,
\Psi_R^{\dagger \bar j} \Psi_L^i\right) \,,
\nonumber\\[3mm]
\{ Q_L^\dagger \,,\,  Q_R \}
&=& -\frac{i}{2\pi}  \int dx\, \partial_x \left( R_{i\bar j}\,
\Psi_L^{\dagger\bar j} \Psi_R^i\right) \,.
\label{twentythree}
\eeqn
This is an anomaly, since the central charge is not seen at the classical level.
Again, this anomaly is in one-to-one correspondence
with a similar anomaly of 4D SYM theory \cite{DvSh}.

The presence of the central charge on the right-hand
side of Eq.~(\ref{twentythree})
implies that there are BPS-saturated kinks interpolating
between the distinct vacua, with the masses 
\beq
M_{\rm fi\,\,  kink} = \frac{1}{2\pi}\left|
\left\langle i\, R_{\bar j\, i}\,
\Psi^{\dagger \bar j}_L \Psi^{i}_R
\right\rangle_{\rm f} -\left\langle i\, R_{\bar j\, i}\,
\Psi^{\dagger\bar j}_L \Psi^{i}_R
\right\rangle_{\rm i}
\right|\,,
\label{fik}
\eeq
where the subscripts i and f mark the initial and final vacua
between which the given kink interpolates. 
These kinks comprise the physical spectrum of the model.
In fact, as is clearly seen from Eqs.~(\ref{fik}) and (\ref{nvalue}),
the kink mass does not depend on specific i and f, but, rather, on the 
difference f--i.

\subsection{$ k$-kinks}

Taking into account Eq.~(\ref{nvalue})
we conclude that the mass of the $k$-kink,
i.e. the kink interpolating between
$|{\rm vac}\rangle_i$ and $|{\rm vac}\rangle_{i+k}$,
is
\beq
M_k = (2\pi)^{-1}\, 2N\Lambda \sin\left(\frac{\pi k}{N}
\right)\,.
\label{kkm}
\eeq
Although the notion of the $k$-kink is self-evident,
a somewhat more specific definition will not hurt.
We will assume   that
the initial and final  vacua are fixed.
For, instance, let us choose as our initial vacuum that with $k=0$
(the right-most point in Fig. ~\ref{ndsv} assuming $\theta =0$).
If the final vacuum has $k= 1$ (the nearest-neighbor vacuum)
then we will refer to the corresponding
interpolation as to the 1-kink. If the final vacuum has $k= -1$,
this is 1-antikink.
If $k=\pm 2$ in the final vacuum
(the next-to-nearest neighbor) we deal with 2-kinks,
and so on.
The interpolation in the anti-clock-wise direction
will be referred to as kink;
the anti-kink interpolates in the clock-wise direction.

The 1-kink is basic, its mass is 
\beq
M_1 =(2\pi)^{-1}\,  2N\Lambda \sin\left(\frac{\pi }{N}
\right) \to \Lambda \,,
\eeq
at $N\to\infty$. The 1-kinks are in one-to-one
correspondence with   the fields
$n^i$ used in Witten's work. The mass  of the $k$-kink
is smaller than $k$ times the mass of the 1-kink, by terms
of the order of $1/N^2$,
\beq
M_k - k M_1 = -\Lambda\,\, \frac{\pi^2}{6N^2}\, k (k^2-1) +O(N^{-4})\,.
\eeq

Thus, the $k$-kinks can be viewed
as bound states of the 1-kinks. Note that these are
{\em not} the bound states discussed by Witten,
as the latter are of different nature
(short-range vs. long-range attractive force). We will return to 
explaining this point later. In Witten's approximation 
only 1-kinks ($n, \, n^*$ fields)  could be seen.

It is instructive to discuss multiplicity of 
$k$-kinks, which will be   helpful in deciding to which  representations  
the $k$-kinks belong.
As we already know, 1-kinks have multiplicity
$N$, they form an $N$-plet.\footnote{This does not include
the supersymmetry degeneracy. Each kink is a member, bosonic or
fermionic, of a short ${\cal N}=2$ supermultiplet, e.g. $n^i$
and $\psi^i$. Remember, that
the final and initial vacua are fixed. The above multiplicities count
only1-kinks which interpolate, say, between $|{\rm vac}\rangle_0$ and 
$|{\rm vac}\rangle_1$ or 2-kinks interpolating 
 between $|{\rm vac}\rangle_0$ and 
$|{\rm vac}\rangle_2$, etc.}
The multiplicities of $k$-kinks for arbitrary $k$ were analyzed in
\cite{Acharya} (see also \cite{Ritz}). If the initial and final vacua i, f are
fixed 
the multiplicity of the $k$-kinks is
\cite{Acharya}
\beq
\nu_k =\frac{N!}{k! (N-k)!}\,.
\eeq
This is consistent with the statement
that $k$-kinks are bound states of $k$ fields $n^i$
of fully antisymmetric type, 
$$n^{[ i_1}n^{i_2} ...  \, n^{i_k]}\,.$$
Thus, if 1-kinks are analogs of fundamental quarks in
QCD, the $k$-kinks emulate $k$-index 
($N$-ality $k$) antisymmetric representation.

In the supersymmetric $CP(N-1)$ model, if one considers
a $k$-kink at the point $x$ and antikink at the point 
$y$ the interaction between them is short-range,
and this pair does not produce a confined system.
As was mentioned, if 
we want to have an analog of the confined quark-antiquark system,
we should depart from supersymmetric limit. For instance,
Witten's work demonstrated that 
1-kink--1-antikink confinement does take place in nonsupersymmetric $CP(N-1)$
model in the large-$N$ limit. 
Simple physics lying behind this phenomenon
will become transparent shortly. Anticipating the result,
I note that the phenomenon occurs because
the vacuum degeneracy is lifted \cite{Markov}. To have a controllable
theoretical framework we must guarantee
the split between the ``former" vacua to be small in the
scale of $\Lambda$. This is easy to achieve provided
the SUSY-breaking mass term (\ref{susybr})
is chosen in such a way that 
$m/g^2\ll\Lambda$.  In what follows
we will limit ourselves to effects of the leading non-trivial
order in 
$m$. In the leading order, i.e. $O(m^0)$,
the result (\ref{kkm}) for the kink masses stays intact.
The kink--antikink tension, which vanishes
in the supersymmetric limit, is generated in the order $O(m^1)$.

\subsection{ Lifting of the vacuum degeneracy and the choice
of vacua}

 At $m \neq 0$
the vacuum degeneracy is lifted. To order 
$m^1$ the vacuum energy density of the $n$-th vacuum becomes
\beq
{\cal E}_n = 2{\rm Re} \left\{\frac{2\, m}{Ng^2} \left\langle i\,  R_{\bar j\, i}\,
\Psi^{\dagger \bar j}_L \Psi^{i}_R
\right\rangle_n \right\} = - \frac{4 m\Lambda }{g^2} \,  \cos\left(\frac{\theta}{N} +\frac{2\pi n}{N}\right)\,.
\label{vacen}
\eeq
To emulate YM theory in a relatively realistic manner
we must make sure that our toy model is (i) $CP$ conserving,  (ii)
the string tension is a $CP$-even quantity and, finally, (iii) the string tension
scales as $N^0$ in the large-$N$ limit.
As has been already mentioned our toy model exhibits
a  variety of dynamical regimes going beyond
the above list. Of this spectrum of dynamical regimes 
I want to choose the one
which meets the requirements (i), (ii) and (iii).

The first requirement implies that $\theta =0$ (in general,  $\pi\times$ integer).
The second requirement implies that for small $\theta \neq 0$
corrections to the $\theta =0$ string tension are $O(\theta^2)$,
rather than $O(\theta )$. This requirement, in conjunction with the third one,
limits possible choices of the initial/final vacua.

Indeed, let us consider the kink interpolating between
$|{\rm vac}\rangle_\ell$ and $|{\rm vac}\rangle_{\ell^\prime}$
with $\ell ,\,\, \ell^\prime = O(1)$. In this 
case, as it follows from
Eq.~(\ref{vacen}), the energy split $$|{\cal E}_\ell -
{\cal E}_{\ell^\prime}| \sim 1/N.$$ We will see
that the interkink tension is proportional
to $|{\cal E}_\ell -
{\cal E}_{\ell^\prime}| $. Thus, this regime is unsuitable
for modeling QCD. This regime was analyzed in Ref.~\cite{5}.
A feeble confinement leads in this case to a nonrelativistic
formula for 1-kink--1-antikink pair,
$$
M_{\rm bound} = 2M_1 + c\, N^{-2/3}\left(m\Lambda\right)^{1/2}\,,
$$
where a constant $c$ depends on the excitation number.
I present this formula here only for the sake of completeness.

\vspace{2mm}

The correct scaling of the tension, $N^0$, is achieved for kinks
interpolating between the
vacua with 
\beq
n = \frac{N}{4} \pm \ell\,, \qquad \ell = O(1)\,.
\label{sk}
\eeq
We will refer to such kinks as symmetric, keeping in mind
the symmetricity of the configuration around $n=N/4$.
In principle, one could consider asymmetric kinks
interpolating between $n=(N/4) +\ell + \ell^\prime$
and $n=(N/4) -\ell +\ell^\prime$ too. In this case the tension,
having the correct
scaling law $N^0$,   violates the requirement (ii).

For the symmetric choice (\ref{sk})
\beq
{\cal E}_{(N/4)+\ell} - {\cal E}_{(N/4)-\ell}
= \frac{8 m\Lambda }{g^2} \,  \sin\left(\frac{2\pi \ell}{N}\right)
\cos\left(\frac{\theta}{N}\right)
= \frac{8 m\Lambda }{g^2} \,  \sin\left(\frac{\pi \, k}{N}\right)
\cos\left(\frac{\theta}{N}\right)
\,,
\label{noc}
\eeq
where $k=2\ell =$ the number of constituents. 
For the asymmetric choice cos$\, \theta /N$ on the right-hand side
would be replaced by cos$\,(2\pi \ell^\prime +\theta )/N$;
its expansion in $\theta$ would contain a term linear in  $\theta$
which would be  inappropriate for emulating QCD (condition (ii)
is not met).

\subsection{ Genuine and metastable vacua}
\label{gmv}

\begin{figure}
 \centerline{\includegraphics[width=2in]{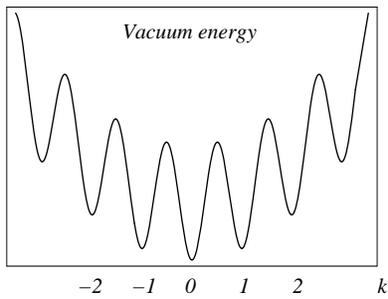}}
\caption{
The vacuum structure in SUSY-broken 
$CP(N-1)$ model at $\theta =0$.}
\label{dva}
\end{figure}

\begin{figure}
 \centerline{\includegraphics[width=2in]{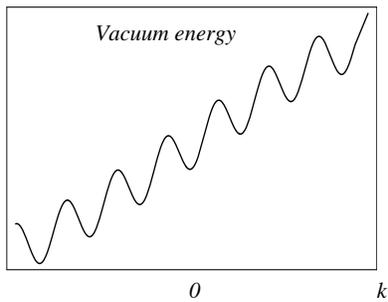}}
\caption{
The vacuum structure in SUSY-broken 
$CP(N-1)$ model at $n=\frac{N}{4} +\ell$ where $\ell= \frac{k}{2} $.}
\label{tri}
\end{figure}

If $n\sim 1$ and $N\to \infty$,
the spacing between metastable vacua
adjacent to the true one ($n=0$) is of the order of
$1/N$ 
as is clearly seen from the expansion
of cosine in Eq.~(\ref{vacen}) (with $\theta =0$)
and the fact that $g^2 \sim N^{-1}$ (the height of
the barriers scales as $N^1$).
This is schematically summarized in Fig. \ref{dva}. 
The probability of the metastable  vacua  decay is proportional
to $\exp (-N\Lambda /m)$,  a straightforward consequence of
 the false vacuum decay theory \cite{Kobzarev:1974cp,Voloshin:1995in}.
The decay probability
vanishes exponentially
at $N\to \infty$, somewhat resembling conventional
(non-SUSY) 4D Yang-Mills theory \cite{nvacym,nvacymp}.
Thus, in this limit each metastable vacuum becomes stable,
not only the one  corresponding to $n=0$.
The fact that the increment of the
vacuum energies scales as $1/N$ in this regime
is responsible for the factor $N$ in the exponent.

Of more interest to us is  the regime (\ref{sk}).
The pattern of ${\cal E}_{(N/4)
+\ell}$ in this regime is
given by 
\beq
{\cal E}_{(N/4) + \ell}  = \frac{4 m\Lambda }{g^2} \,  \sin\left(\frac{2\pi \ell }{N}\right)
\label{vacenp}
\eeq
 (at $\theta =0$)
and schematically depicted in Fig.~\ref{tri}. In this case the point $\ell =0$ does not
correspond to the absolutely stable minimum; rather,  the $\ell =0$ state
is metastable, as well as all neighboring minima lying in the vicinity 
of $\ell =0$, say, $\ell =\pm 1, \,\pm 2,$
etc. The increment of the vacuum energy density
in the subsequent metastable minima is $O(N^0\, m\Lambda)$.
The suppression factor  in the probability of the metastable 
vacuum decays becomes
$$N^{-1} \exp (-\Lambda /m)\,.$$ 
The large-$N$ suppression shows up only in the pre-exponent.
Not to break the applicability of the approximation used, 
we need to keep the SUSY-breaking parameter
$m$ small, i.e.  $\Lambda /m \gg 1$. This ensures that
the suppressing exponent is operative, and 
the metastable vacua under consideration are in fact stable.
Since $m$ is in our hands, this is always doable.
It is of paramount importance that the increment of the
vacuum energy 
in the subsequent minima is $O(N^0)$ in this regime.
This translates in the statement $\sigma = O(N^0)$.

\subsection{  $k$-kink tension}

\begin{figure}
 \centerline{\includegraphics[width=3in]{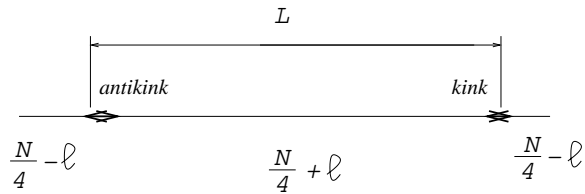}}
\caption{
A $k$-kink separated from $k$-antikink by distance $L$.}
\label{chet}
\end{figure}

Now we are fully prepared to consider 
$k$-kink--$k$-antikink confinement in a QCD-like regime.
Let us consider a  $k$-kink interpolating between
$|{\rm vac}\rangle_{(N/4)-\ell}$ and $|{\rm vac}\rangle_{(N/4)+\ell}$,
with $k=2\ell$ on the right, with   the corresponding antikink
on the left.
Schematically this configuration is depicted in Fig. \ref{chet}.

The vacuum energy density 
in the interval between the kinks is higher than that
outside,
\beq
\Delta {\cal E} 
= \frac{8 m\Lambda }{g^2} \,  \sin\left(\frac{\pi \, k}{N}\right)
\,.
\eeq
At large $L$ the overall energy of the configuration
depicted in Fig. \ref{chet} behaves as 
\beq
\Delta {\cal E} \, L + 2 M_k\,.
\eeq
From this we conclude that  the tension of the string confining the $k$-kinks is
\beq
\sigma_k=\frac{8 m\Lambda }{g^2} \,  \sin\left(\frac{\pi \, k}{N}\right).
\label{fte}
\eeq
At large $N$
this tension is $N$ independent (remember, $g^{-2}\sim N$), just 
like it is
$N$ independent in QCD. One should remember that the combination
$m/g^2$ is renormalization-group invariant.

Using Eq.~(\ref{kkm}) one can rewrite the same expression
as 
\beq
\sigma_k = (8\pi m  /Ng^2)M_k\,.
\eeq
Since $m$ is arbitrarily small,
confinement is weak (but not suppressed by $1/N$), 
and the $k$-kinks at the string ends 
can be viewed as static sources, analogs
of the probe QCD quarks in the $k$-index 
antisymmetric representation of color.
It is remarkable that the $k$-kink tension
follows the exact sine formula!

I hasten to add a few words about a limitation of this parallel, the
presence of two dimensionful parameters,
$m$ and $\Lambda$, as opposed to the only parameter $\Lambda$
of YM theory. To keep  valid approximations vital for our consideration,
one must insist that $m \ll \Lambda$. As supersymmetry breaking
increases, and $m$ approaches $\Lambda$,
theoretical control erodes, and is finally
lost at $m \sim\Lambda$. 

One can ask what happens if one
interchanges the position of the $k$-kinks in Fig. 
\ref{chet}. This would amount to the substitution
$\ell\to -\ell$, resulting --- formally ---
 in the sign change of the interkink tension
(\ref{fte})! In other words, confinement seems
to give place to anticonfinement (linear repulsion).
This is the consequence of the fact that the vacua we work with
are metastable rather than stable. The latter circumstance was imposed on us: in order to emulate $N$-independent string tension of QCD
we had no other choice in the toy model at hand.
Two kinks repelling each other in fact describe
the process of relaxation of an excited string into a less
excited state through   production of a pair of lumps.  I 
should remind
that these are the ``wrong" excited strings, with no analogs in QCD, see
the end of Sect.~\ref{naswc}.

\subsection{ Lessons for QCD?}

In two-dimensional theory $k$-kinks are confined much in the
same way as $k$-quarks in QCD. If we choose a QCD-like regime, with the
tension $O(N^0)$, the  ``$k$-string" spectrum follows the sign formula.
This statement is valid if and only if
$m\ll\Lambda$, i.e. supersymmetry breaking is small. (One cannot put $m=0$, however,
since in the supersymmetric limit kinks in CP$(N-1)$ model
are not confined at all).

This may be the most important lesson.
Other analyses which produced the sign formula
in four dimensions  either 
rely on supersymmetry directly or
also assume that its breaking is somehow suppressed. 
One may conjecture
that the sine law for the $k$-string tension
takes place  in large-$N$ QCD approximately. A hidden small numerical parameter
justifying the approximation may
emerge due to a ``residual supersymmetry" of QCD.
This is not a slip of the tongue. One can argue  \cite{rsusy}
that pure gluodynamics has something like rudimentary supersymmetry!
Elaboration of the nature and accuracy of this approximation remains 
an open task.

\section*{Acknowledgments}

I am grateful to Adi Armoni, A. Gorsky, Igor Klebanov,  and Alyosha Yung
for valuable comments.
Special thanks go to Philippe de Forcrand for providing me with data on which
Sect.~\ref{ndol} is based.
This work was supported in part by DOE grant DE-FG02-94ER408.


\end{document}